\documentstyle[psfig]{article}
\begin{document}
\begin{titlepage}
\begin{flushright}
KEK-CP-045 \\
April 1996 
\end{flushright}
\vspace*{10mm}
\Large
\begin{center}
STRINGS IN COMPUTER\footnote[1]{Based on a contribution paper for 
the proceedings of the international conference ``Frontiers in 
Quantum Field Theory'' celebrating Professor Kikkawa's 60th 
birthday, held at Osaka University, Osaka, Japan, December 
14-17, 1995. \\
Presented by T.Yukawa.}
\end{center}
\vspace{10mm}
\large
\begin{center}
H.Kawai\footnote[2]{E-mail address: kawaih@theory.kek.jp}, 
N.Tsuda\footnote[3]{A JSPS Research Fellow, E-mail address: ntsuda@theory.kek.jp}, 
and 
T.Yukawa\footnote[4]{E-mail address: yukawa@theory.kek.jp}
\end{center}

\normalsize

\begin{center}
$^{\dag}$ $^{\ddag}$
National Laboratory for High Energy Physics (KEK) \\
       Tsukuba, Ibaraki 305 , Japan \\
\vspace{0.5cm}
$^{\S}$
Coordination Center for Research and Education, \\
The Graduate University for Advanced Studies, \\
Hayama-cho, Miura-gun, Kanagawa 240-01, Japan \\
and \\
National Laboratory for High Energy Physics (KEK) \\
Tsukuba, Ibaraki 305, Japan
\end{center}
\vspace{1cm}
\parindent 5cm
\begin{abstract}
Complex structures are determined for surfaces with $S^2$ and $T^2$ topologies
generated by the dynamical triangulation method. For a surface with $S^2$ 
topology the spacial distribution of the conformal mode is obtained, while for 
the case of $T^2$ topology the distribution of the moduli parameter is 
calculated. 
It is also shown that the network of Feynman diagrams of massive $ \phi ^3 $ 
scalar theory has a unique complex structure.
This gives a numerical justification of the hadronic string model for 
explaining the n-particle dual amplitude.
\end{abstract}
\end{titlepage}


\section{Complex structure in the Polyakov string}

For quantizing the Polyakov string one usually considers a set of closed 
string world sheets of the 2-dimensional Euclidean space-time\cite{PO}. 
Each world sheet is considered to be a closed and orientable manifold with 
various topology. 
Once the Riemannian metric $g_{ab}$ is given on this manifold, it is always 
possible to introduce a complex structure on the surface. 
For example, the Riemannian metric of a surface with $ S^2 $ topology can be 
written  by a suitable choice of local coordinates as 
$$ g = g_{ab} dx^a dx^b = e^{\sigma (z)} dz d{\bar z} ,$$
where $ z=x + {\rm i} y $, and $ e^{\sigma (z)} $ is the conformal factor.
Therefore the problem of quantizing two-dimensional gravity is reduced to 
considering the quantum fluctuations of the conformal mode and the complex 
structure. 
This is the basic assumption one usually employs in the Liouville field theory 
of non-critical Polyakov string.

Now, we would like to examine this assumption constructively, by
determining the complex structure for 
those surfaces obtained numerically by the dynamical triangulation(DT). 
The Monte Carlo simulation of the random surface by the 
DT method\cite{DT} has been regarded as a numerical 
exercise of the matrix model\cite{matrix_model} which is known to give the 
correct critical exponent identical to the continuous field theory, {\it i.e.} 
the Liouville field theory\cite{Liouville}. 
In what follows we shall show that the DT method also exhibits its explicit 
correspondence to the continuous field theory in the sense that those surfaces 
created by DT numerically can be decomposed to a unique complex structure and 
conformal mode\cite{Kaw_Tsu_Yuk}.

\section{The basic idea  -resistivity-}

Our method is based on the observation that the resistance of a homogeneous 
conducting sheet is invariant under the local scale transformation. 
This can be shown by considering the resistance of a small rectangular 
section with length $ a $ and width $ b $ of a conducting sheet. 
The resistance between two sides of length $ a $ is given by 
$$ R = r { a \over b} ,$$
where $ r $ is the resistivity constant. The resistance $ R $ is apparently 
invariant under the scale change { $a \mapsto a \eta$}, 
{$b \mapsto b \eta$}.
Therefore, the local scale factor does not affect the resistance.

\begin{figure}
\vspace*{0cm}
\centerline{
\psfig{file=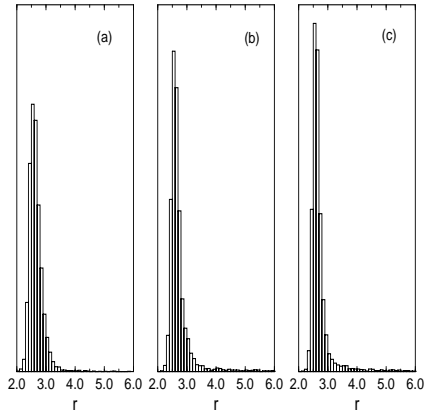,height=6cm,width=6cm}}
\vspace{0cm}
\caption
{
Resistivities for the case of pure gravity. 
The size of ($a$), ($b$) and ($c$) are $4000$, $8000$ and $16000$ triangles 
respectively.
}
\label{fig:Pure_Resist}
\vspace{0cm}
\end{figure}

\begin{figure}
\vspace*{0cm}
\centerline{
\psfig{file=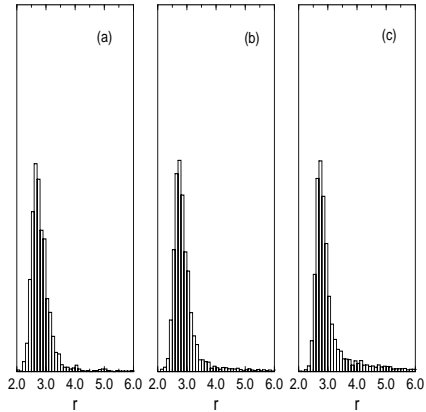,height=6cm,width=6cm}} 
\vspace{0cm}
\caption
{
Resistivities for the case of 2 scalars.
}
\label{fig:B2_Resist}
\vspace{0cm}
\end{figure}

Experimentally, the resistance measurement of a conducting sheet is carried 
out by picking up four points on the surface which we specify by four  
complex numbers $ \{ z_i \} $. With a source of current $ I $  placed at 
$ z_3 $ and a sink of the current at $ z_4 $ the voltage drop from a point 
$ z_1 $ to a point $ z_2 $ is written as 
$$ V(z_1)-V(z_2) = R(z_1,z_2;z_3,z_4) I .$$
In the case of surface with $S^2$ topology it can be regarded as an infinite 
flat sheet and  we obtain from Gauss' law
$$ R = -{r \over 2 \pi } \ln | [z_1,z_2;z_3,z_4]| ,$$
where 
$$[z_1,z_2;z_3,z_4] = {z_1-z_3 \over z_2-z_3}{z_2-z_4 \over z_1-z_4}$$
is known as the anharmonic ratio. 

We regard the dual graph of a triangulated surface as a trivalent network of 
registers each having $ 1 \Omega $. We pick up two vertices for applying current 
with $ 1A $, and solve the Kirchhoff equation by the Jacobi iteration method. 
Then, we measure voltage drops between other pair of vertices. Using the 
$ SL(2,C) $ transformation, 
$$ z \mapsto {a z + b \over c z + d}    (a d - b c = 1 ), $$ 
invariance property of the anharmonic ratio we can fix three points among 
$ \{ z_i \} $ at $ z_3 = 0 $,$ z_4 = \infty $ and $ z_2 = 1 $ without changing 
the resistance by an 
appropriate choice of four complex parameters $(a,b,c,d)$. 
In practice, among $ 4! $ combinations of measuring 
arrangements there are known to be only two independent measurements, 
while we have three unknowns, namely one complex $ z_1 $ and one real $ r $. 
In order to fix them unambiguously we add one more measuring point $ z_5 $.   

\begin{figure}
\vspace*{-1.5cm}
\centerline{\psfig{file=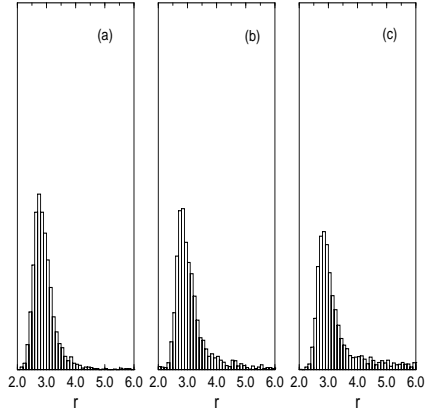,height=6cm,width=6cm}} 
\vspace{-0.4cm}
\caption
{
Resistivities for the case of 3 scalars.
}
\label{fig:B3_Resist}
\vspace{0cm}
\end{figure}

\section{Complex structure of  $ S^2 $ surfaces}

Let us show  results of the resistivity measurement for random surfaces 
with $S^2$ topology simulated by the DT method with $ N $ triangles. 
If a conducting surface is uniform and homogeneous, we should get a fixed 
resistivity regardless of points of measurements.  
Fig.1 shows the distributions of resistivity constant for surfaces of three 
area sizes with no matter fields coupled(the pure gravity). 
The distributions peak at about $ r=2.6 $ and they get sharper as the size 
grows.  Broader distributions in small surface simulations 
reflect the mesh structures 
of circuit networks, and we can expect that the peak will eventually become 
the $ \delta$-function type distribution as the area get infinity.  
On the contrary, when surfaces are coupled with matter fields with the 
central charge $c$ bigger than 1, peaks of the resistivity distribution 
tend broader as the size grows. 

\begin{figure}
\vspace{-0.2cm}
\centerline{
\psfig{file=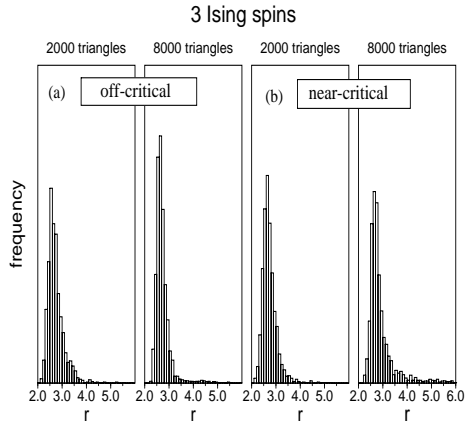,height=6cm,width=6cm}} 
\vspace{-0.9cm}
\caption
{
We plot resistivities for 3 Ising spins both case of off-critical $(a)$ and 
near critical $(b)$.
}
\label{fig:Ising3}
\vspace{0cm}
\end{figure}

Fig.2($ c=2 $) and Fig.3($ c=3 $) show the tendency clearly. 
This phenomenon suggests the absence of a unique continuous limit for cases 
$ c>1 $ as expected from analytic theories.  
The $ c=1 $ barrier can be seen more significantly in the measurement of 
a surface with three Ising spins coupled(Fig.4). 
It should behave as a system with  the $c={3 \over 2}$ matter coupled at 
the critical point, while it is expected to be in the same universality 
class as the pure gravity off the critical point. 
Indeed, our measurement shows this behavior. Broadening of the 
resistivity for bigger size naturally implies the transition of
the surface to the branched polymer phase beyond $ c=1 $. 
  
\section{Distribution of the conformal factors}

Once we obtain the resistivity $ r $ for a surface, we can locate the position 
of each triangle by two measurements of resistances, for example
$$ R( z_1,1;0,\infty) = -{r \over 2\pi} \ln | z_1 | ,$$
and
$$ R(z_1,0;1,\infty) = -{r \over 2\pi} \ln | 1-z_1 | .$$
By sweeping all the rest of triangles with fixing three points which 
correspond to $ (1,0,\infty) $, we can determine the position of all the 
triangles of this surface. The point density obviously represents the 
distribution of conformal factors proportional to
$$\lim_{z_{4} \to \infty} |z_{4}|^{4} <e^{\alpha \sigma(0)} e^{\alpha 
\sigma(1)} e^{\alpha \sigma(z_{4})} e^{\alpha \sigma(z_1)}> .$$
In Fig.5 we show the distribution of triangles whose position $ z $ is 
mapped within a circle by the transformation
$$ z \mapsto {az \over z+a-1} $$
with $ a = {1 \over 2} + i {{\sqrt 3} \over 2}$. 
\begin{figure}
\vspace*{0cm}
\centerline{\psfig{file=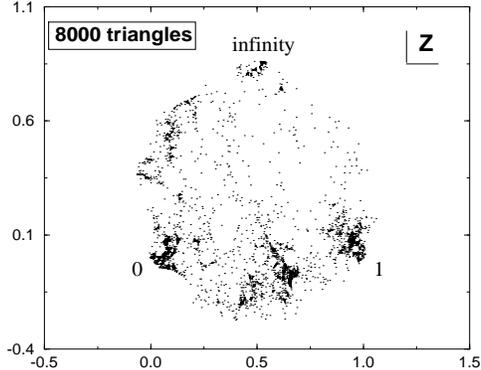,height=6cm,width=7.5cm}} 
\vspace{-0.5cm}
\caption[]
{
A plot of the obtained coordinates of all vertices for $c=0$ with five 
independent configurations superimposed.
As a matter of convenience all points are mapped into a circle with the 
center at ($\frac{1}{2},\frac{\sqrt{3} }{2}$) and radius $\frac{\sqrt{3}}{3}$.
}
\label{fig:Cplx_Coord}
\vspace{0cm}
\end{figure}
There exists a theoretical prediction of the distribution\cite{Zamo}, but 
it has not been presented in a form ready for the direct comparison to the 
numerical experiments.

\section{Complex structure of the $ T^2 $ surface}

As far as two dimensional orientable manifolds of $ S^2 $ topology are 
concerned they all have the same complex structure, i.e. any manifold can 
be transformed to the other by a combination of a general coordinate 
transformation({\it Diffeo}) and the Weyl transformation({\it Weyl}). 
On the other hand for the surface with $ T^2 $ topology the moduli space
$$\{ g_{ab} \} / Weyl \otimes Dif \! feo$$
is spanned by a complex plane of moduli $ \tau $. 
It is defined by the ratio of the integrals of an Abelian differential 
$$ \tau = { \oint_b j_\mu d x^\mu + i \oint_b {\tilde j}_\mu d x^\mu \over
         \oint_a j_\mu d x^\mu + i \oint_a {\tilde j}_\mu d x^\mu} ,$$
where integration contours (we call them as the a-cycle and the b-cycle) 
are chosen to be two closed paths on the manifold which cross each other 
only once. 
Here, $j_{\mu} dx^{\mu}$ is a harmonic $1$-form and $\tilde{j}_{\mu} dx^{\mu}$ 
is its dual.
If we regard $j_{\mu}$ as a current density on the torus, 
$r \oint_{a} j_{\mu} d x^{\mu}$ and $\oint_{a} {\tilde j}_{\mu} d x^{\mu}$ 
correspond to the voltage drop along the a-cycle and the total current 
crossing the a-cycle, respectively.

In practice, we select two closed paths made up by connecting edges of 
triangles, which intersect only once. Then we cut the surface 
along one of the path for
which we choose the b-cycle, and apply constant voltages(1V) in between 
neighbouring triangles dual to the cut(Fig.6).
\begin{figure}
\vspace*{0cm}
\centerline{\psfig{file=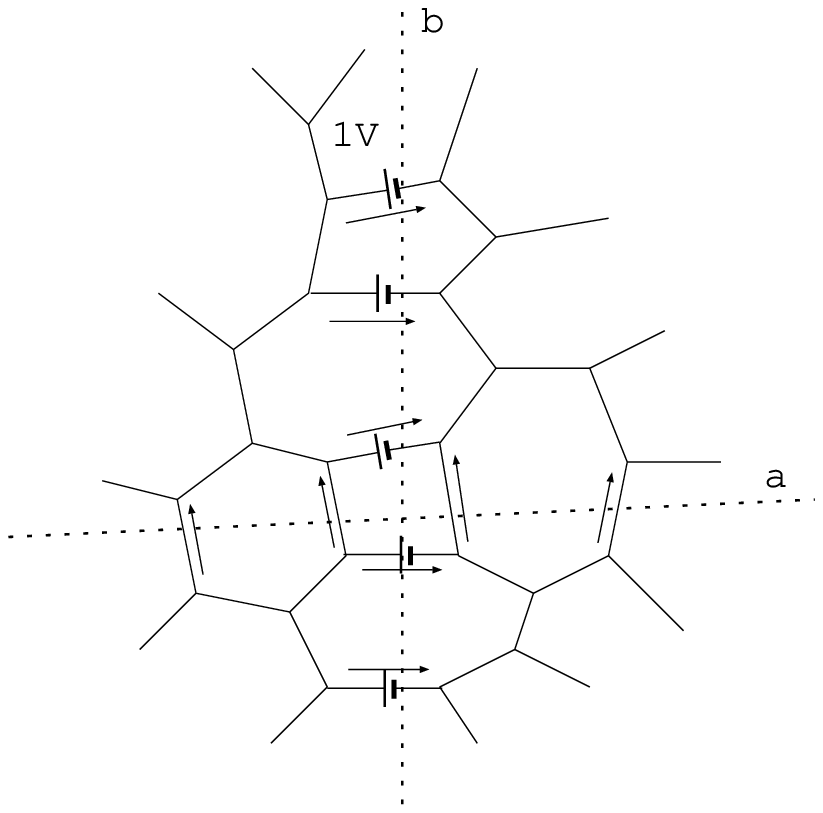,height=6cm,width=6cm}} 
\vspace{0cm}
\caption{}
\label{fig:net_current}
\vspace{0cm}
\end{figure}
Solving the Kirchhoff equation we can determine all the currents $ j_\mu $ 
and $ {\tilde j}_\mu $ along links between neighbouring triangles in the 
dual graph. 
By construction of the network the following two integrals are trivial;
$$ \oint_a j_\mu d x^\mu = {1 \over r} ,   \oint_b j_\mu d x^\mu = 0 .$$
Thus $\tau$ is obtained by measuring total currents crossing two cycles;
$$
\tau = \frac{i I_{b}}{\frac{1}{r} + i I_{a}}
$$
where $I_{a}$ and $I_{b}$ represent the total currents 
crossing the a-cycle and the b-cycle, respectively. 
Instead of measuring  the resistivity of this surface we employ the value 
determined previously for the surface with $S^{2}$ topology and 
corresponding central charge. 

\begin{figure}
\vspace*{-1.0cm}
\centerline{\psfig{file=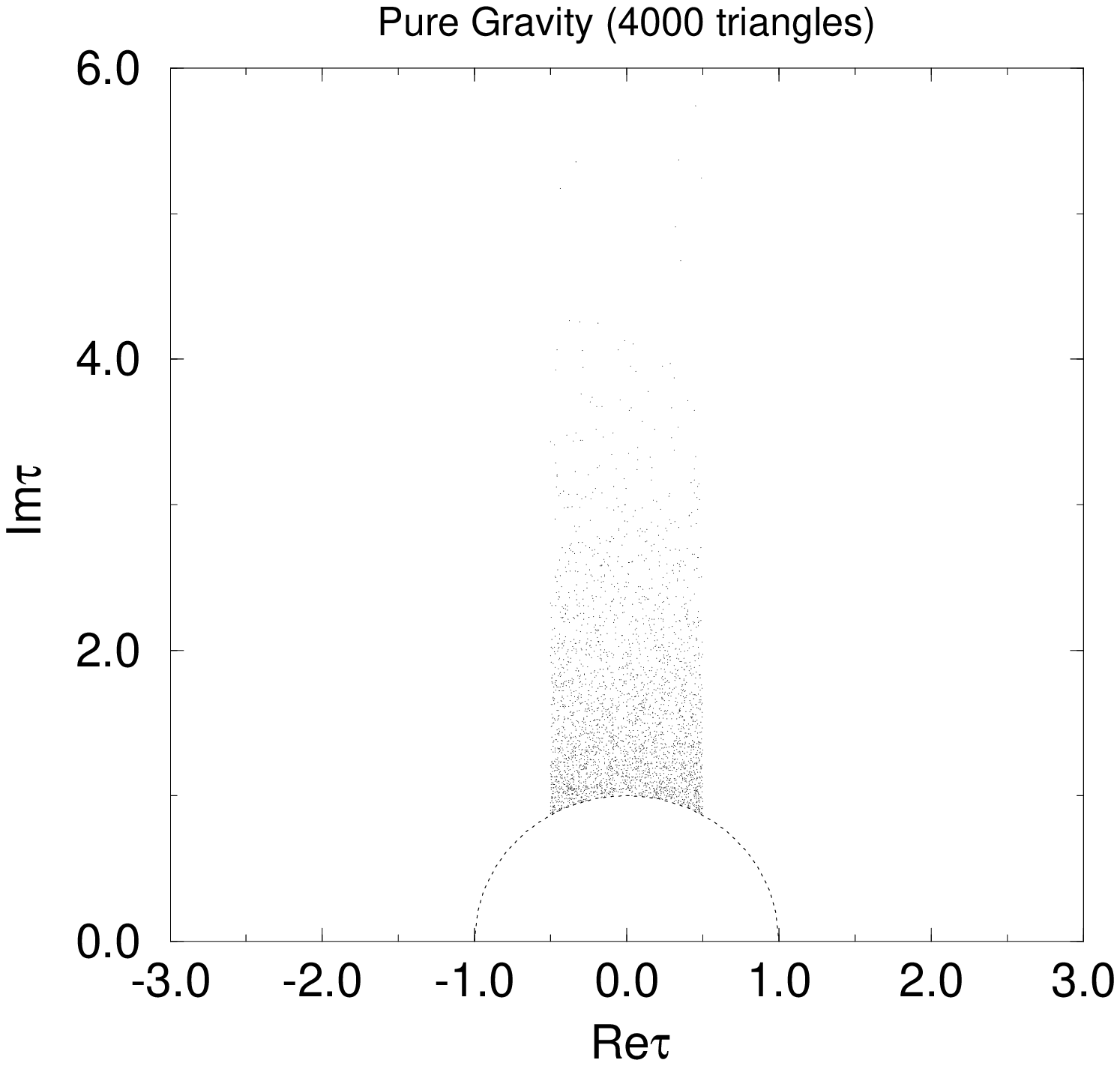,height=7cm,width=9cm}} 
\centerline{\psfig{file=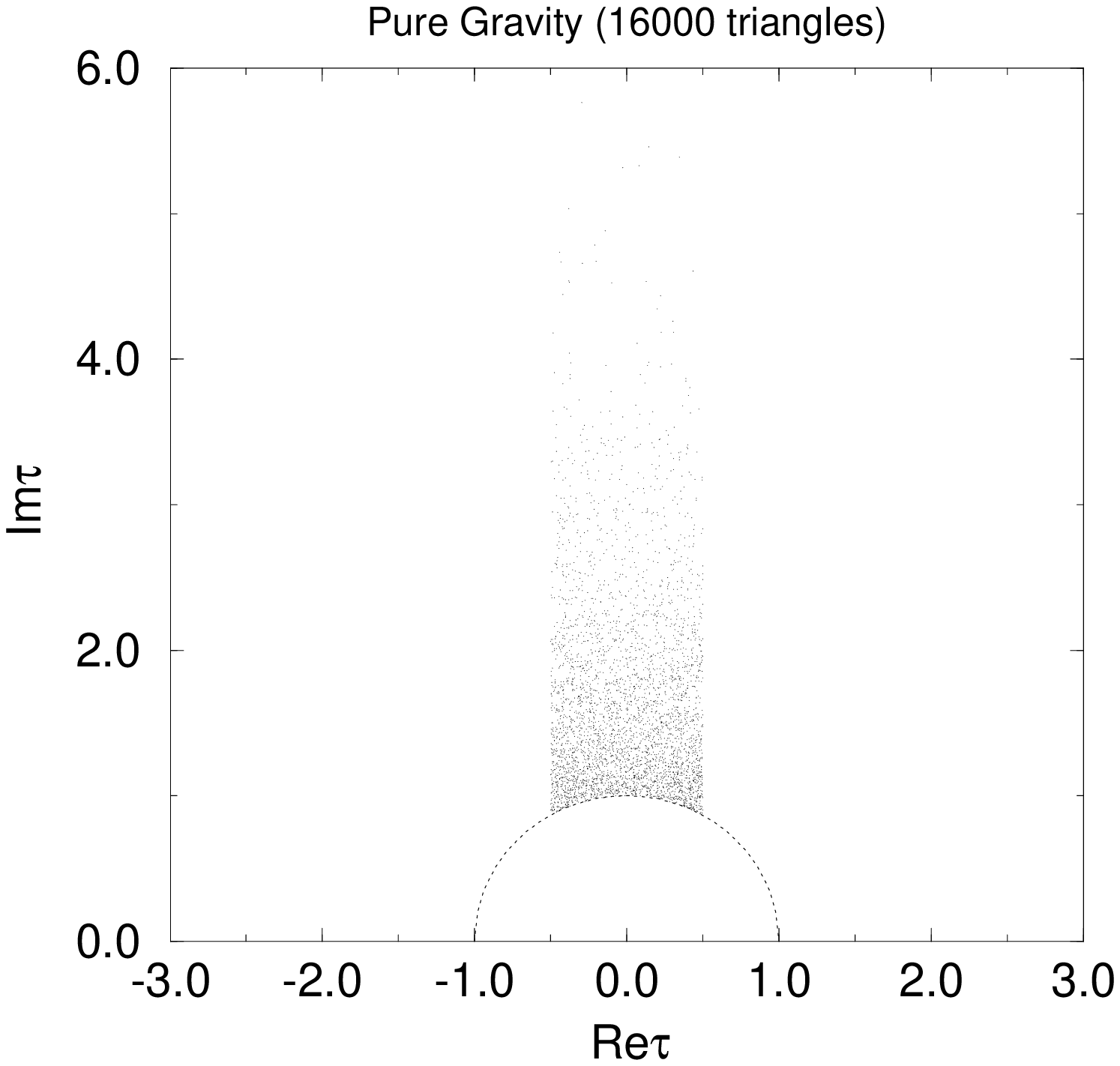,height=7cm,width=9cm}} 
\vspace{-0.5cm}
\caption
{
Plot of the moduli($\tau$) on the complex-plane mapped to the 
fundamental domain with a total number of triangles of $4,000$ and $8,000$.
Each dot corresponds to the moduli$\tau$ for a sample of DT-surface.
}
\label{fig:Moduli}
\vspace{0cm}
\end{figure}
Fig.7 shows the distribution of $\tau$ for configurations of 4,000 and 
8,000 triangles. Each value of $\tau$ is transformed appropriately by the 
$ SL(2,Z) $ transformation 
$$
\tau \mapsto {a\tau+b \over c\tau+d} \;\;\; (ad-bc=1)
$$
so that it is in the fundamental domain. By summing over real $\tau$'s 
with a fixed imaginary part we get the distribution for surfaces 
with two, four and eight thousand triangles in Fig.8, which should be compared 
with the theoretical prediction
$$
({\rm Im} \tau)^{-{3 \over 2}} e^{-{\pi \over 6} {\rm Im} \tau} 
\left| \prod_{n=1}^{\infty} (1- e^{2\pi i \tau n}) \right|^{-2}.
$$
\begin{figure}
\vspace*{0cm}
\centerline{\psfig{file=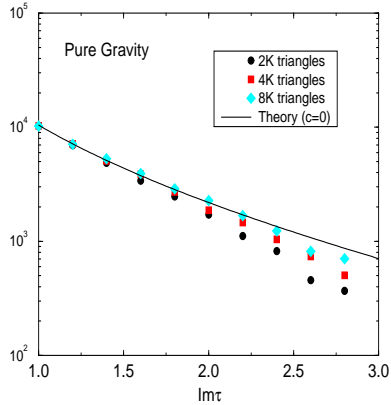,height=6cm,width=6cm}} 
\vspace{0cm}
\caption
{
Four density-distributions(histogram) of imaginary part of moduli 
$\tau_{2}$ for the case of pure-gravity with the topology of the torus.
These are obtained by integrating over of the real part of moduli $\tau_{1}$
on the complex plane.
A scale of vertical line is arbitrary.
}
\label{fig:Hist_Comp_Pure}
\vspace{0cm}
\end{figure}
The Liouville theory predicts for a surface with c scalar fields
$$
({\rm Im} \tau)^{-{c+3 \over 2}} e^{{\pi \over 6}(c-1) {\rm Im} \tau}
\left|\prod_{n=1}^{\infty} (1- e^{2\pi i \tau n}) \right|^{-2(c-1)},
$$
{\it i.e.} for simulations with $c>1$ matter the surface will be 
unstable to the polymer like torus configuration in the limit 
${\rm Im} \tau \rightarrow \infty$.

\section{Hadronic strings -a root of the string model-}

\begin{figure}
\vspace*{0cm}
\centerline{\psfig{file=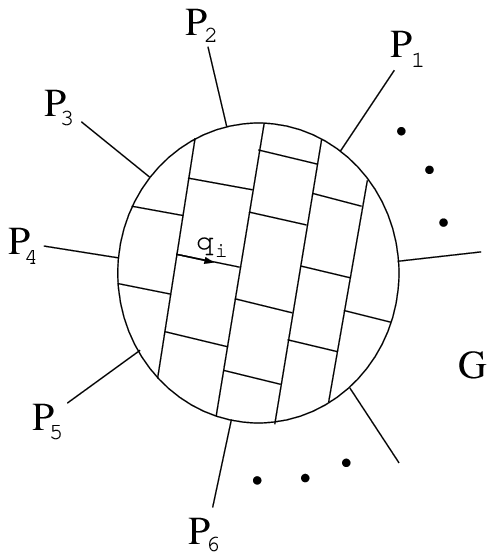,height=7cm,width=6.5cm}} 
\vspace{0cm}
\caption{}
\label{fig:net_graph}
\vspace{0cm}
\end{figure}
The n-particle amplitudes of the massive $\phi^3$ scalar theory 
corresponding to the diagram $G$ shown by Fig.9 is given by
$$I_G(p_1,p_2,...,p_n) =\int \prod^{N_1}_{i=1} d^{D}q_i \prod^{N_1}_i
{1 \over q^2_i+m^2}\prod^{N_0}_{a=1}\delta(P_a) ,$$
where $P_a$ is the net momentum flowing into the vertex $a$. Here, $N_0$ 
stands for the total number of vertices, and $N_1$ for the total number of 
internal lines. By the Laplace transform of a propagator,  
$${1 \over q^2_i + m^2} = \int^{\infty}_0 d\alpha_i e^{-\alpha_i (q_i^2 
+ m^2)} ,$$ 
the amplitude is expressed after integrated over $\{q_i\}$ as 
$$I_G=\int \prod_i^{H_1} d\alpha_i \alpha_i^{-{D \over 2}} 
 e^{-\sum \alpha_i m^2}\int \prod_a^{N_0} d^Dx_a {\rm exp}\{ -{1 \over 4}\sum_i 
{(x_{a(i)}-x_{b(i)})^2 \over \alpha_i}\} e^{i\sum_a^{N_0}x_a p_a} ,$$
where $a(i)$ and $b(i)$ are two vertices connected by the propagator $i$. 
This equation has been often noticed on its resemblance to the electric 
circuit\cite{fishnet_analog},{\it i.e.}
$$
\alpha_i \rightarrow {\rm resister},  x_a \rightarrow {\rm voltage}, 
p_a \rightarrow {\rm current}.
$$
Then, $Q ={1 \over 4}\sum {(x_{a(i)}-x_{b(i)})^2 \over \alpha_i}$ is regarded 
as the total heat generated by the circuit.  It has been conjectured that as the 
mesh becomes finer the surface will tend to be a uniform, homogeneous 
and continuous conductor. In this case after averaging over all n-particle 
diagrams we will obtain
$$<I> \sim e^{-{1 \over 2}\sum_{a,b}R_{ab}({\bar \alpha} ) p_a p_b} ,$$
where ${\bar \alpha}$ is an appropriate resistivity of the surface and the 
resistance is same as that of the flat conductor;
$$ R_{ab} \propto \ln | z_a-z_b |$$.

We have made the measurement of resistivity for  networks with $S^2$ topology 
constructed by the same DT method as described previously. This time each 
link has resistance distributed randomly with the probability
$$\alpha^{-{D \over 2}} e^{-\alpha m^2} .$$
Here, we have chosen the parameters to be $D=0$ and $m=1$ in the simulation. 
\begin{figure}
\vspace*{0cm}
\centerline{\psfig{file=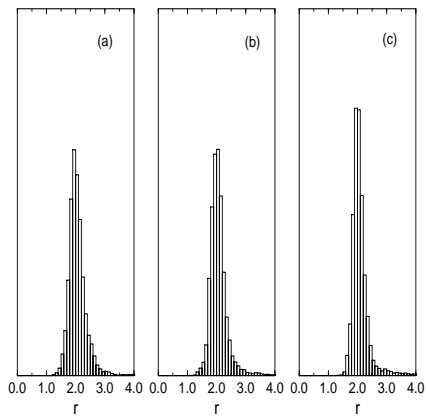,height=6cm,width=6cm}} 
\vspace{0cm}
\caption
{
Resistivities for the case of pure gravity when each resistance is randomly 
distributed.
}
\label{fig:Random_resist}
\vspace{0cm}
\end{figure}
Fig.10 shows the size dependence of the resistivity distribution. 
As the size gets larger the distribution becomes sharper similar to the case 
of the pure gravity. 
This gives a numerical support of the theoretical conjecture.

\section*{Acknowledgments} 
The authors wish to give heartly thanks to professor Kikkawa for his guidance 
to the physics of quantum string, which he has given us at various occasions, 
such as {\it the summer school for young physicists} and the lecture series 
at KEK. His lectures have been always lively and stimulating. 
We also acknowledge Ishibasi, HariDass and members of the KEK theory
group for discussions and comments.

\end{document}